\documentclass[referee]{mn2e}

\input psfig.sty

\def\T1{\ {$T_1$}\ }
\def\MT1{\ {$M_{T_1}$}\ }
\def\ct1{\ {$(C-T_1)$}\ }
\def\CT10{\ {$(C-T_1)_0$}\ }

\def\VI0{\ {$(V-I)_0$}\ }

\def\2cd{\ {two-color diagram}\ }

\def\sf{\ {specific frequency}\ }
\def\ell{\ {elliptical}\ }

\def\gtsim{\ {\raise-0.5ex\hbox{$\buildrel>\over\sim$}}\ }
\def\ltsim{\ {\raise-0.5ex\hbox{$\buildrel<\over\sim$}}\ }

\begin{document}

\title[LMC star clusters]{The star cluster frequency throughout the Large Magellanic Cloud}

\author[A.E. Piatti]{Andr\'es E. Piatti\thanks{E-mail: 
andres@oac.uncor.edu}\\
Observatorio Astron\'omico, Universidad Nacional de C\'ordoba, Laprida 854, 5000, C\'ordoba, 
Argentina\\
}

\maketitle

\begin{abstract}
We address the issue about the 
variation of the star cluster frequency (CF) in the Large Magellanic Cloud (LMC) in terms of the cluster 
spatial distribution. We adopted the
LMC regions traced by Harris \& Zaritsky \shortcite{hz09} and used an updated version of the cluster 
database compiled by Baumgardt et al. \shortcite{betal13}. The CFs were produced by taking into 
account an appropriate selection of age bins. 
Since the uncertainty in a cluster's age 
can be large compared to the size of the age bins, we account for the fact that a
cluster could actually reside in one of a few adjacent age bins. 
We confirm that there exist 
some variations of the LMC CFs in terms of their spatial distributions, although some caveats should be
pointed out. 30 Doradus resulted to be the region with the highest relative frequency of youngest
clusters, while the log($t$) = 9-9.5 (1-3 Gyr) age range is featured by cluster 
formation at a higher rate 
in the inner regions than in the outer ones. We compared the observed CFs to theoretical 
CFs, which are based on the star formation histories of the field stars in each region
of the LMC, and found the former predicting more or fewer clusters than observed depending on the
field and age range considered. 
\end{abstract}


\section{Introduction}

The star cluster frequency (CF) - the number of clusters per time unit as a function of age -
in the Large Magellanic Cloud (LMC) has been the subject of recent studies by Baumgardt et al.
\shortcite[hereafter BPAG]{betal13} and de Grijs et al. \shortcite[hereafter dGGA]{dgetal13}.
Both works considered the LMC cluster population as a whole and produced
overall CFs for the galaxy. BPAG showed that about 90$\%$ of all clusters older than 200 Myr are 
lost per dex of lifetime, which implies a cluster
dissolution rate significantly faster than that based on analytic estimates and N-body simulations. 
However, dGGA showed that there is no evidence of significant destruction, other than
that expected from stellar dynamics and evolution in simple population models for ages up to 
1 Gyr (log($t$)=9).

On the other hand, different studies show that the LMC field star formation rate has varied from 
one place to another in the galaxy \cite{cetal05,hz09,cetal11,retal11}. Particularly, Harris \&
Zaritsky (hereafter HZ) concluded from the concordance between the star formation and chemical 
enrichment histories of the field and cluster populations that the field and cluster star 
formation modes are tightly coupled. Carrera et al. \shortcite{cetal11} found that 10 different
studied fields have age-metallicity relationships (AMRs) statistically distinguishable and that
the disc AMR is similar to that of the clusters and is well reproduced by closed-box models or
models with a small degree of outflow. Recently, Piatti \& Geisler \shortcite{pg13} studied 21
LMC fields spread across the main body of the galaxy and found that the cluster and field AMRs 
show a satisfactory match only for the last 3 Gyr (log($t$)=9.5), while for the oldest ages 
 ($>$11 Gyr (log($t$)= 10.05)) the
cluster AMR is a remarkable lower envelope to the field AMR. 

From these results, it would appear reasonable to infer that if the star formation history (SFH) of 
LMC  field stars has been different throughout the galaxy, and  field stars and clusters show
some evidence of sharing the pace of their formation and chemical enrichment, then the CF should 
reflect the same spatial variation as seen in the field stars. However, as mentioned above, the LMC CF 
has been studied without making distinction on the cluster position. Thus, it is feasible that regions
with a noticeable recent star formation activity were treated together with those more quiescent 
outer regions, and consequently the resultant CF pictured a mixed behaviour of the individual
CFs. Here, we attempt to build CFs for different LMC regions taking advantage of those regions
delimited by HZ and compare them to each other in order to find out some hints for any dependence
of the CF with the position in the galaxy. Fortunately, we have also the chance of comparing our
results to the CFs coming from theoretical models computed from the SFHs recovered by HZ for 
different LMC regions.

The paper is organised as follows: Sect. 2 is devoted to describe the assembly of the available
cluster age estimates into a robust compilation and to  establish some procedures in order to properly 
deal with age bins and errors. Then, in Sect. 3 we discuss the level of completeness of the
presently available cluster sample with age estimates and adopt some statistical rules for
a reliable analysis of the resultant CFs, which is performed in Sect. 4 along with the comparison
with theoretical CFs. Finally, Sect. 5 summarises the main conclusions.

\section{Handling of the star cluster sample}

We are  interested in building the CFs for clusters located in the HZ regions
(see their Fig. 6), namely: the Bar, the Outer Bar, 30 Doradus, the Southeast Arm, the
Northwest Arm, the Blue arm, Constellation III, and the Northwest Void. We realise that
there are other parts of the LMC that could be meaningful from a particular 
astrophysical point of view. However, we prefer to use the HZ regions since 
they correspond to different galactic substructures  where we can investigate CF 
variations. Furthermore, their available SFHs are useful to compute CFs from
theoretical models as well.  Recently, van der Marel \& Kallivayalil \shortcite{vdmk13}
showed that for most tracers (including clusters and field stars), the LMC's line-of-sight
velocity dispersion is at least a factor $\sim$2 smaller than their rotation
velocity, which implies that the whole LMC is a kinematically cold disc system.  
Therefore, we assume that there probably has not been a statistically 
significant mixing of clusters from one region to another.     

We used the catalogue of LMC clusters kindly provided by H. Baumgardt which contains ages, 
luminosities and masses for 1649 clusters (BPAG). We made an update of its ages and
included more than 50$\%$ of the BPAG's intermediate-age clusters from our own catalogue of LMC
clusters with ages and metallicties put into an homogeneous scale 
\cite{getal97,betal98,petal99,petal02,getal03,petal03a,petal03b,petal09,petal11a,p11,p12,p13,petal13},
as well as all the clusters with ages estimated by Pietrzynski \& Udalski \shortcite{pu00}. We have 
paid attention  to completing as much as possible  of the sample of clusters older than 1 Gyr
(log($t$)=9) since we are interesting in building CFs covering most of the galaxy lifetime. We do not focus on
its youngest end, for which we refer to dGGA. Fig. 1 depicts all the LMC clusters with age
estimates represented by open circles. In the figure, we also included the LMC cluster candidates
compiled in the general catalogue of extended objects by Bica et al. 
\shortcite[hereafter BBDS]{betal08} represented by dots. Table 1 lists the number of clusters in the 
BBDS's catalogue with age estimates (open circles with central dots in fig. 1), as well as
those with ages that were identified in later  catalogues (taken from BPAG, open
circles without central dots in Fig. 1). Notice that most of the clusters with age estimates have 
been included in the BBDS's compilation. 

The first premise in building the CFs consisted in considering the age errors. Indeed, by taking into 
account such 
errors, the interpretation of the resultant CFs can differ appreciably from that obtained using 
only the measured ages without accounting for their uncertainties. However, the treatment of age 
errors in the CFs is not a straightforward task. Moreover, even if errors did not play an important 
role, the binning of age ranges could also bias the results. For instance, Bonatto et al. 
\shortcite{betal06} and Wu et al. \shortcite{wetal09} used different fixed age intervals to build 
cluster age distributions using the same cluster database and found remarkably  different
results. At first glance, a fixed age bin size is not appropriate for yielding the intrinsic age 
distribution, since the result depends on the chosen age interval and the age errors are typically a 
strong function of the age. On the contrary, an age bin whose width is of the order of the age errors
of the clusters in that interval appears to be more meaningful. This would lead to the selection of 
very narrow bins (in linear age) for young clusters and relatively broader age bins for the older 
ones.

Piatti \shortcite{p10} took the uncertainties in the age estimates of Galactic open clusters
into account in order to define the age intervals, with the aim of building an age histogram that 
best reproduces the intrinsic age  distribution. 
Indeed, the age errors 
for very young clusters are a couple of Myrs, while those for the oldest clusters are at least a few 
Gyrs. Therefore, smaller bins are appropriate for young clusters, whereas larger bins are more 
suitable for the old clusters. In practice, he varied the bin size based on the average error of the 
age of the clusters that fall in each bin.
Piatti et al. \shortcite{petal11a,petal11b} have also used these
precepts for producing age distributions of clusters older than 1 Gyr (log($t$)=9)
in both Magellanic Clouds.

In order to  account for the effect of the age uncertainties in the CFs, we searched our list of 
LMC clusters and found that typical age errors are between 0.10 $\la$ $\Delta$log($t$) $\la$ 0.15. 
Therefore, aiming at tracing the variation of the age uncertainties along the whole age range, we set 
the age bin sizes according to this logarithmic law to build CFs for the different HZ regions. We used
intervals of $\Delta$log($t$) = 0.10. At the same time, we focused on an additional issue. Even though
the age bins are set to match the age errors, any individual point in the CF  may fall in the 
respective age bin or in any of the two adjacent bins. This happens when an age point does not fall in
the bin centre and, due to its errors, has the chance to fall outside it. Note that, since we chose 
bin dimensions as large as the involved errors, such points should not fall on average far beyond the 
adjacent bins. However, this does not necessarily happen to all age points, and we should consider at 
the same time any other possibility. 

For our purposes, we first considered the cluster age range split in bins with sizes
following the logarithmic law mentioned above. On the other hand, each age point with its error 
($\sigma(age)$) 
covers a segment whose size is given by 2$\times$$\sigma(age)$, and may or may not fall centred on 
one of the age bins, and has dimensions smaller, similar or larger than the age bin wherein it is 
placed. These scenarios generate a variety of possibilities, in the sense that the age segment could 
cover from one up to 5 age bins depending on its position and size. For this reason, we weighed the 
contribution of each age point to each one of the age bins occupied by it, so that the sum of all the 
weights equals unity. The assigned weight was computed as the fraction of its age segment
[2$\times$$\sigma(age)$] that falls in the age bin. In practice, we focused on a single age
bin and computed the weighted contribution of all the age points to that age bin. Then we repeated the
calculation for all the age bins. The challenge of knowing whether a portion of an age point (an age 
segment strictly speaking) falls in an age bin, was solved by taking into account the following 
possibilities of combination between them.  For each age interval we looked for clusters with ages 
that fall inside the considered age bin, as well as clusters where age +/- $\sigma(age)$
could cause them to fall in the considered age bin.  Note that if age 
+/- $\sigma(age)$ causes a cluster to step over the considered age bin
(e.g. from one bin younger to one bin older), then we consider that that
cluster may have an age that places it inside the considered age
interval as well.

To illustrate how CF built with and without bin size and error effects differ, we plot 
in Fig. 2 the CF obtained by BPAG and that from our approach represented by filled circles 
and the solid line, respectively. Both CFs have been normalized to the total number of clusters.

\section{Completeness of the star cluster sample}

BPAG and dGGA have imposed mass limits to their LMC cluster samples in order to deal with
statistically complete lists. BPAG built the global LMC CF with 322 clusters older than 10$^7$ yrs, 
brighter than $M_V$ = -3.5, and more massive than 5000 M$_{\odot}$, from an original list of 1649 
clusters. This means that they used the 322 clusters as representative of the LMC cluster population 
rather than the extended 1649 cluster sample. On the other hand, dGGA conducted different analyses by
using a 50$\%$ completeness limit, $M_V$ = -4.3 mag, based on single stellar population models. In
this case, the minimum mass varies as a function of the age and the CF was consequently built from
different cluster subsamples.

As a starting point, we also followed the recommendation of managing a statistically significant sample 
of clusters by applying some sort of selection.  Since we basically used the 
mass values provided by H. Baumgardt for the BPAG cluster list, and those obtained by Piatti 
\shortcite{p11} for the additional intermediate-age clusters of our own catalogs,
we constrained the cluster sample to those with masses higher than  5$\times$10$^3$ M$_{\odot}$ and 
(1.8$\times$log($t$) - 12.8) M$_{\odot}$ for ages younger and older than 1 Gyr (log($t$)=9), respectively.
The resultant CFs normalized to the total number of clusters used for each HZ region are shown in 
Fig. 3 with dashed lines. Notice that they account for bin size and age uncertainty effects, so
that they show the main fiducial features of the LMC CFs.

Bearing in mind that we are interested in examining possible variations in the relative shape of the
CFs in terms of the cluster position in the galaxy, some unavoidable questions arise: what is the
benefit of using the above statistically complete sample instead of the whole available cluster
sample? How meaningful are the results based on the statistical sample?, etc. In order to bring over an
answer, we built CFs including all the available clusters in the different HZ regions. As for the
statistically constrained sample, they are also corrected from bin size and age error effects and
normalized to the total number of clusters used. Fig. 3 depict the resultant CFs with solid lines.
 As can be seen, the BPAG's global mass cut-off does not result in a satisfactory
statistically representative sample of the cluster population in the different HZ regions. 

In order to find a more appropriate mass cut-off for the individual HZ regions, we produced mass
distributions using age intervals of $\Delta$log($t$) = 0.2, from log($t$) = 7.0 to 10.0, and took the
lower mass values of the full width at half maximum (mean mass - $\sigma$(mass)) of those observed 
cluster mass distributions. The 84$\%$ more massive (encompassed in the 
[mean mass - $\sigma$(mass), highest mass] interval) of the whole cluster sample reached in general
masses lower than 10$^3$ M$_{\odot}$ for ages younger than 1 Gyr (log($t$)=9). Then, we produced normalized CFs 
using this new limit (see Fig. 4.)
%
The difference between normalized CFs based on the statistically limited sample and on the whole
cluster sample is mostly negligible, with the exception of some excess in the latter of 
intermediate-age clusters and very young clusters in the Blue Arm and in the Northwest Void,
respectively. From this result we conclude that this mass cut-off sample is statistically
representative of the whole sample of clusters with age estimates. Furthermore, Table 1 shows the number
of clusters used in the mass cut-off sample and the percentage that they represent respect to the total 
number of clusters with age estimates. As can be seen, such percentages range from 28$\%$ up to 63$\%$, 
with an average of 47$\%$. This means that nearly the 50$\%$ more massive of the whole cluster sample 
with age estimates mostly trace the overall behaviour of the CFs.  Alternatively, clusters with masses 
below the mass cut-off limit do not contribute significantly to the normalized CFs, although they do 
represent on average nearly 50$\%$ of the total number of clusters with age estimates. 

According to Fig. 1, the spatial distribution of catalogued clusters without age estimates (dots not
encircled) does not seem to be particularly remarkable throughout the HZ regions. They are more or
less distributed between those with age estimates (dots encircled and open circles) covering similar
areas. One exception arises: the Southeast Arm. For the remaining HZ regions, the number of catalogued 
clusters without age estimate represent a relative minority respect to the total number of catalogued 
clusters as listed in Table 1 (column 5). In this context,  we wonder whether the clusters without 
ages can be assumed to follow the same age distribution as the clusters with ages. We address this
issue by assuming nine hypothetical different age distributions for the total number of clusters without
age estimates. We cover scenarios from placing the clusters in the same age bin until 
distributing them uniformly along the age range log($t$) = 7.0 - 9.0. Then, we added these age 
distributions to those  previously obtained for the different HZ regions, and computed the respective 
normalized CFs. The resultant CFs (see Appendix A for details) show that, for most of the HZ regions,
clusters without ages could 
affect the CFs only if all of them fell in the youngest or in the oldest age bin (log($t$) = 7.0 or 9.0, 
respectively). In the case of the Southeast Arm, any proposed scenario remarkably affects the CF.


An additional issue we would like to address is related to the completeness of the LMC cluster 
catalogue with respect to the whole LMC cluster population. The BBDS's catalogue include objects discovered
by the $HST$, by the Optical Gravitational Lens Experiment (OGLE) \cite{u03}, etc. While the latter 
covers the central regions of the galaxy, the former only spots isolated small fields. On the other 
hand, the Magellanic Cloud Photometric Survey (MCPS) \cite{zetal04} used by Glatt et al. 
\shortcite{getal10} to compile their own cluster database reaches a limiting magnitude between $V$ = 
20 and $V$ = 21 mag, depending on the local degree of crowding in the images \cite{netal09}. According
to dGGA, the depth of the observations made by OGLE is of the order 1.5 mag shallower than the MCPS. 
This means that the BBDS's catalogue is not as deep as that of Glatt et al. Indeed, some clusters
with age estimates in the BPAG sample -who used the Glatt et al.'s compilation- are not included in 
the BBSD's catalogue (see open circles in Fig. 1). Nevertheless, since the derived CFs do not 
seem to show any arguable effect due to catalogued clusters without age estimates (except for the
Southeast Arm), it might be also reasonable that our possible incompleteness of non-catalogued clusters 
(supposed to be fainter than the catalogued ones) do not play a role in the CFs either.


\section{The star cluster frequencies}

Fig. 5 depicts the resultant CFs from the statistically constrained cluster sample, corrected by
bin size and age uncertainty effects. We have selected 
those objects which match the mass cut-off requirements mentioned above. Since all the CFs have been 
normalized to the total number of clusters employed, we did not need to shift them by any constant 
value for comparison purposes. Indeed, they are simply superimposed. From the figure, we confirm that 
there exist some variations of the LMC CF in terms of their spatial distributions, although some 
caveats should be pointed out. For instance, it seems that the period during which most of the CFs 
resemble  each other occurred from log($t$) $\sim$ 8.0 to 8.4. However, more recently (log($t$) 
$<$ 7.5) differential
 cluster formation rates have taken place, 30 Doradus being the region with the highest relative 
frequency of youngest star clusters in the galaxy. The Northwest Void presents the lowest CF during 
the last 30 Myr; the Outer Bar is at an intermediate level between 30 Doradus and the Northwest Void,
while the Bar, the Southeast Arm, the Northeast Arm, and the Blue Arm have had a relatively lower
cluster formation activity than the Outer Bar.

We also see interesting results in the log($t$) $\sim$ 9-9.5 (1-3 Gyr) range,
a period during which the Magellanic Clouds may have tidally interacted
with each other (e.g. Diaz \& Bekki 2011, Besla et al. 2012) and show an increase in the star 
formation rate of field stars \cite{wetal13}. Besides
the fact that this period of cluster formation is associated to bursting formation events after
the enigmatic cluster age gap \cite{p11,pg13}, the resultant CFs show that such a formation period 
was more intense in the Bar, the Outer Bar, and the Northwest Arm; of an intermediate  strength in 
30 Doradus and the Northwest Void; while Constellation III, the Southeast Arm, and the Blue Arm
account for the lowest cluster formation level. Focusing on the spatial distribution of these
regions, the above result might be pointing out that the reservoirs of gas out
of which the clusters were formed at that time were more important in the inner regions 
than in the outer ones, with some exception. Finally, the oldest LMC clusters located within
the 8 HZ regions mainly populate the Bar, the Northwest Arm, and the Blue Arm. As is well-known, 
their origin is still a conundrum. While some authors suggest that they were formed during a very early 
and rapid period of enrichment \cite{pg13},  others showed that some of those found in the outskirts
of the LMC could have originally belonged to the Small Magellanic Cloud \cite{cetal13}. Any way, 
our results witness that the oldest star clusters are not isotropically distributed.

Finally, we compared the present CFs to those obtained from theoretical models. The models were 
kindly provided by H. Baumgardt (personal communication). They assume that clusters are born with a 
power-law mass distribution with slope $\alpha$ = -2 and with a rate that is proportional to the field 
star formation rate determined by HZ for each individual field, respectively, along with their 
corresponding uncertainties. The models then apply cluster dissolution due to stellar evolution, 
two-body relaxation and an external tidal field according to the EMACSS code \cite{ag12,getal13}.
In addition, he used 300.000 clusters for each HZ region in order to have a good statistical sample. 
We applied mass cut-offs as described in Sect. 3 and normalized the theoretical CFs by the total
number of clusters used, so that they can directly be compared to the observed ones. Fig. 6 shows the 
observed CFs (thin solid line) with the theoretical ones superimposed (thick gray solid line); the 
uncertainties in the latter being drawn with thick gray dotted lines. As can be seen,
while the shapes of the theoretical CFs generally follow those of the observed CFs, the theoretical CFs 
vary  between predicting more or fewer clusters than observed depending on the
field and age range considered.
If cluster dissolution were relatively well-known and easily modeled, then the above results
would lead us to conclude that the LMC cluster population has not evolved as a coupled or 
independent system to the field star population, exclusively, but as a combination of both scenarios 
that likely have varied in importance in different regions during the lifetime of the galaxy, with 
the coupled mode being dominating.

However,  Lada \& Lada \shortcite{ll03} suggested that most, if not all, stars form in some
sort of cluster.  This implies that field stars are the result of cluster dissolution and not from an 
independent formation mechanism. Based on their results, any difference between the observed and
theoretical CFs would then be the result of different cluster dissolution rates as a function of mass 
and/or region.  To this respect, there has been considerable debate about cluster dissolution rates and 
whether the rate is dependent on mass and environment, or if a certain percentage of clusters are 
destroyed each year regardless of any cluster or field properties (see, e.g., Bastian et al. 2011, 
and references therein). In this context, Fig. 6 would lead us to conclude that cluster
dissolution rates are not universal, but instead may be affected by the
masses of the clusters formed or the different environments in the LMC.

During the most recent star formation epoch (log($t$) $<$ 30 Myr), the observed CFs in the Blue Arm, 
the Northwest Void, and possibly the 
Northwest Arm are lower than the corresponding counterparts in the theoretical scenario.
However,  there is a role reversal towards older ages until log($t$) $\sim$ 8.5.
On the contrary, the field star formation processes would seem to be relatively less significant
 than those for the clusters in the Bar and the Outer Bar during the last 30 Myrs. In the cases
of the Outer Bar, 30 Doradus, and Constellation III such a trend keeps up until log($t$) $\sim$ 8.5.
As the log($t$)= 9-9.5 (1-3 Gyr) age range is concerned, the theoretical CFs obtained for the Blue 
Arm, Constellation III, Southeast Arm, and less importantly for the remaining HZ regions, appear to
be higher than the observed ones. This behaviour might be originated by either a simple more active
field star formation rate or by cluster dissolution or by both effects combined. Nevertheless, notice
that for the outermost HZ regions not only the theoretical CFs (or indirectly the field star 
formation rates) are higher, but also the observed CFs are lower than those for the remaining HZ
regions.

\section{Final Remarks}

In this work we address for the first time the issue of the variation of the LMC CF in terms of the 
cluster spatial distribution. For this purpose, we adopted the LMC regions traced by HZ, namely: the 
Bar, the Outer Bar, 30 Doradus, the Southeast Arm, the Northwest Arm, the Blue arm, Constellation 
III, and the Northwest Void. As for the cluster database, we used that of BPAG, which was updated by 
including more than 50$\%$ of their intermediate-age clusters as well as those from OGLE. 

When building the CFs we took into account the influence of adopting arbitrary age bins, as well as
the fact that each age value is associated to an uncertainty which allows the age value to fall 
centred on an age bin or outside it. We employed a procedure which achieves a compromise between the 
age bin size and the age errors.
Particularly, we adopted an age bin size which varies with the age as a 
logarithmic law ($\Delta$log($t$) = 0.1). Then, we considered the possibility that the extension 
covered by an age value (properly a segment of 2$\times$$\sigma$(age) long) may have a dimension smaller, 
similar or larger than the age bin wherein it is placed. The assigned weight  was computed as 
the fraction of its age segment that falls in the age bin.

For each HZ region we produced mass distributions using age intervals of $\Delta$log($t$) = 0.2, from
log($t$) = 7.0 to 10.0, and took the lower mass values of the full width at half maximum (mean mass - 
$\sigma$(mass)) of those observed cluster mass distributions. The 84$\%$ more massive 
of the whole cluster sample - encompassed in the (mean mass - $\sigma$(mass), highest mass)
interval - reaches  in general masses lower than 10$^3$ M$_{\odot}$ for ages younger than 1 Gyr 
(log($t$)=9).
In order to deal with a 
statistically  significant cluster sample, we constrained the cluster sample to those with masses 
higher than 10$^3$ M$_{\odot}$ and (1.8$\times$log($t$) - 12.8) M$_{\odot}$ for ages younger and older
than 1 Gyr  (log($t$)=9), respectively. Such mass cut-off sample includes from 28$\%$ up to 63$\%$, with an average 
of 47$\%$, of the more massive clusters in the whole cluster sample.

We confirm that there exist some variations of the LMC CFs in terms of their spatial distributions, 
although some caveats should be pointed out. For instance, it seems that the period during which most 
of the CFs resemble one to each other occurred from log($t$) $\sim$ 8.0 to 8.4. However, more recently 
(log($t$) $<$ 7.5) differential cluster formation rates have taken place, 30 Doradus being the region 
with the highest relative frequency of young star clusters in the galaxy, while the Northwest Void 
presents the lowest CF during the last 30 Myr; the remaining HZ regions having intermediate levels of
cluster formation activity. During the  log($t$)=9-9.5 (1-3 Gyr) age range the resultant CFs show that the cluster
formation proceeded more intensely in the inner regions than in the outer ones, while the oldest
LMC clusters located within the 8 HZ regions  mainly populate the Bar, the Northwest Arm, and the 
Blue Arm.

Finally, we compared  the present CFs to those obtained from theoretical models assuming cluster 
formation rates similar to the star formation rates determined by HZ for their individual LMC regions. 
We found that while the shapes of the theoretical CFs
generally follow those of the observed CFs, the theoretical CFs vary   
between predicting more or fewer clusters than observed depending on the
field and age range considered.

\section*{Acknowledgements}

We are grateful for the comments and suggestions raised by the anonymous
reviewer which helped us to improve the manuscript.
We thank Holger Baumgardt for providing us with the theoretical models as well as
with constructive suggestions.
This work was partially supported by the Argentinian institution
Agencia Nacional de Promoci\'on Cient\'{\i}fica y Tecnol\'ogica (ANPCyT).

\clearpage

\setcounter{table}{0}
\begin{table}
\caption{LMC cluster statistics.}
\begin{tabular}{@{}lcccc}\hline
HZ region   & \multicolumn{2}{c}{Clusters with age estimates} &  Clusters in the & Clusters without\\
            &  in BBDS & in other catalogs & mass cut-off sample & age estimates\\\hline

Bar               &  347 & 60 &  206 (51$\%$)& 52 (11$\%$) \\ 
Outer Bar         &  168 & 43 &  114 (54$\%$)& 29  (12$\%$) \\
30 Doradus        &  65  & 63 & 80  (63$\%$)& 19  (12$\%$) \\
Southeast Arm     &  47  & 14 & 17  (28$\%$)& 50  (45$\%$) \\
Northwest Arm     &  69  & 19 & 47  (54$\%$)& 17  (16$\%$) \\
Blue Arm          &  84  & 63 & 73  (50$\%$)& 11  (7$\%$) \\
Constellation III &  60  & 55 & 53  (46$\%$)& 2   (2$\%$) \\ 
Northwest Void    &  52  & 24 &  24 (32$\%$)& 25  (25$\%$)\\

\hline
\end{tabular}
\end{table}

\begin{figure}
\centerline{\psfig{figure=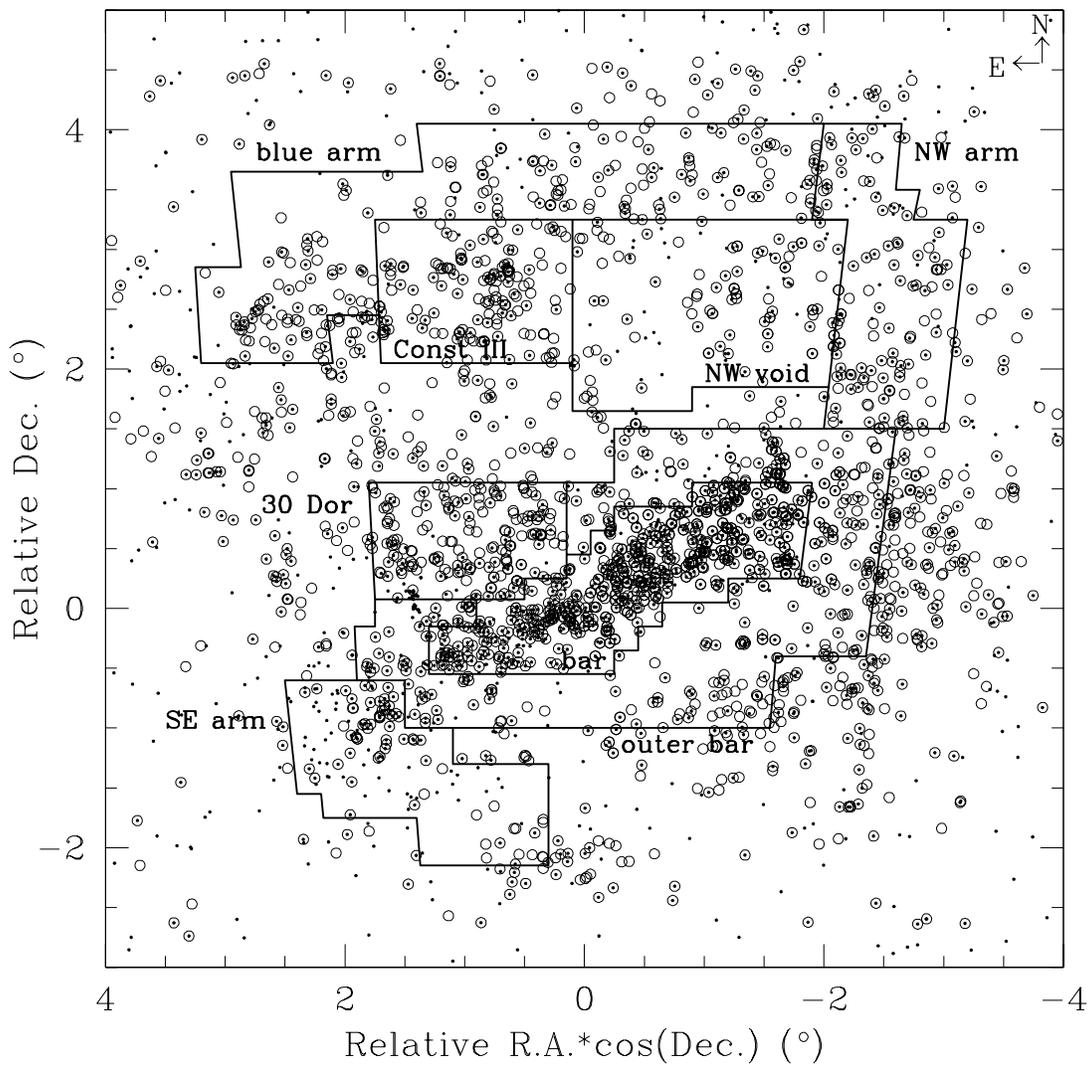,width=144mm}}
\caption{Spatial distribution of LMC clusters: those with an open circle have
age estimates available. The clusters included in the BBDS's catalogue are
represented by dots. The HZ regions are also overpotted.}
\label{fig1}
\end{figure}

\begin{figure}
\centerline{\psfig{figure=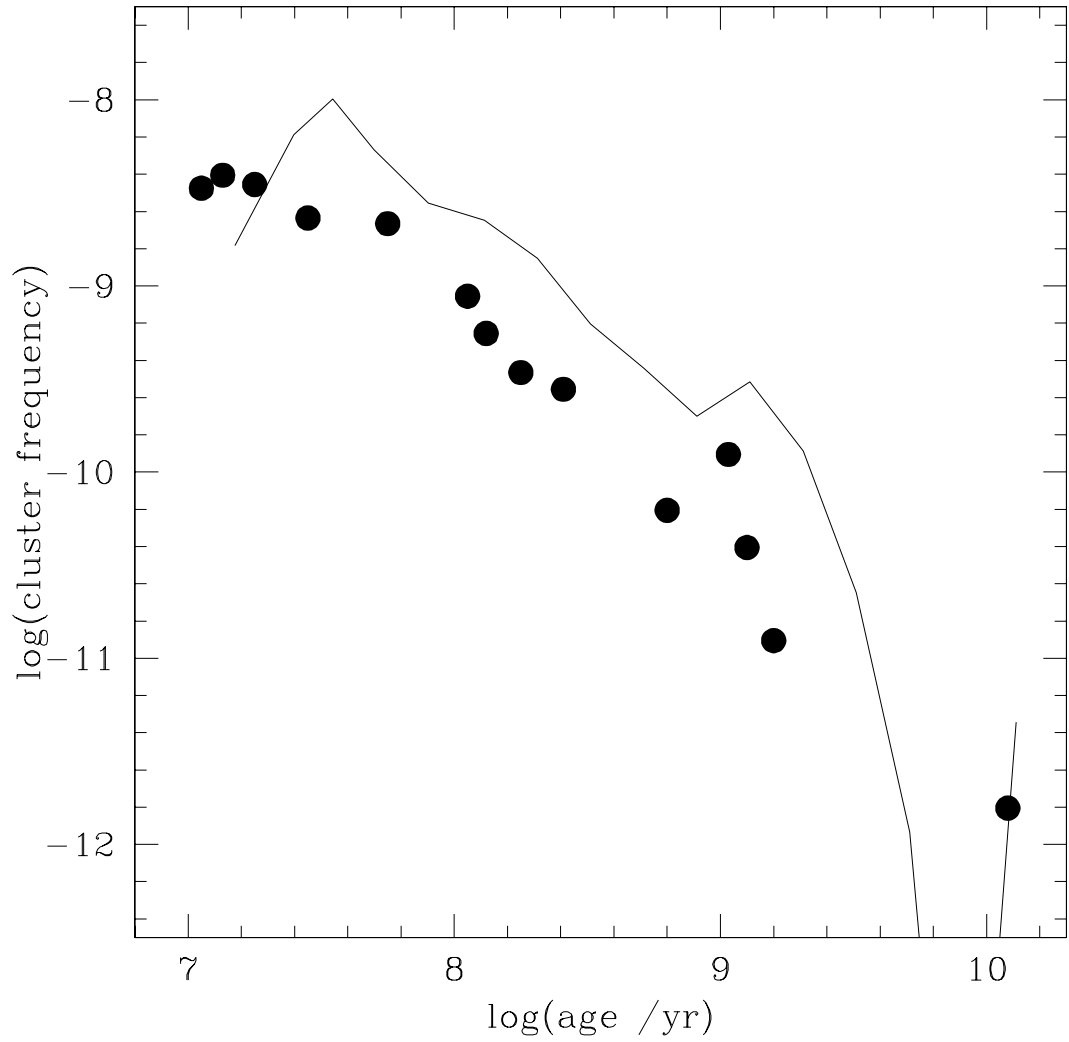,width=144mm}}
\caption{CFs obtained by BPAG and in the present work as described in Sect. 2 for the
322 BPAG's cluster sample, represented by filled circles and by a solid line, respectively.
Both CFs have been normalized to the total number of clusters.}
\label{fig2}
\end{figure}


\begin{figure}
\centerline{\psfig{figure=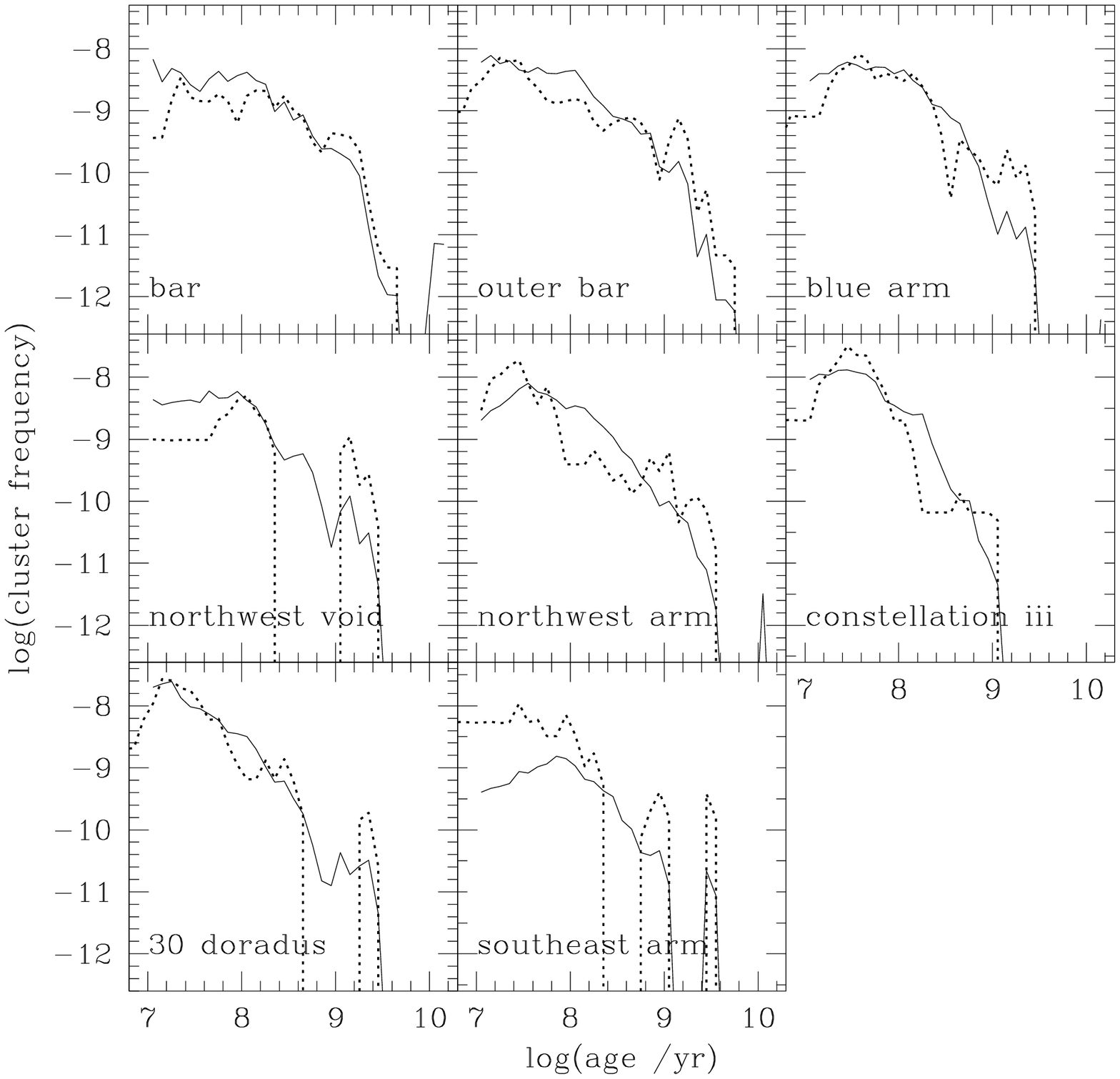,width=144mm}}
\caption{5$\times$10$^3$  M$_{\odot}$ cut-off CFs (dashed lines) compared to those built using the whole cluster sample
(solid lines).}
\label{fig3}
\end{figure}

\begin{figure}
\centerline{\psfig{figure=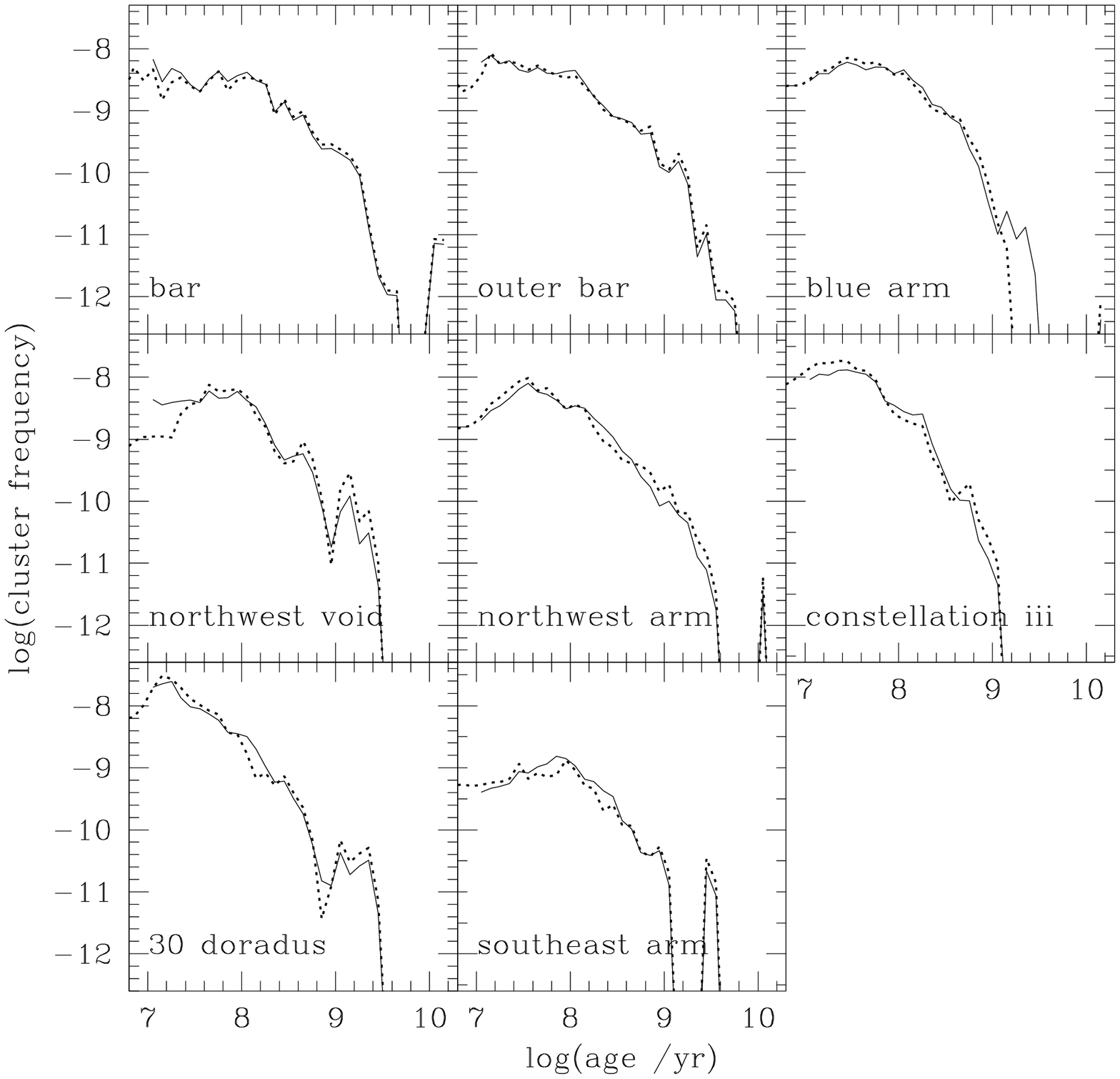,width=144mm}}
\caption{10$^3$  M$_{\odot}$  cut-off CFs (dashed lines) compared to those built using the whole cluster sample
(solid lines).}
\label{fig4}
\end{figure}

\begin{figure}
\centerline{\psfig{figure=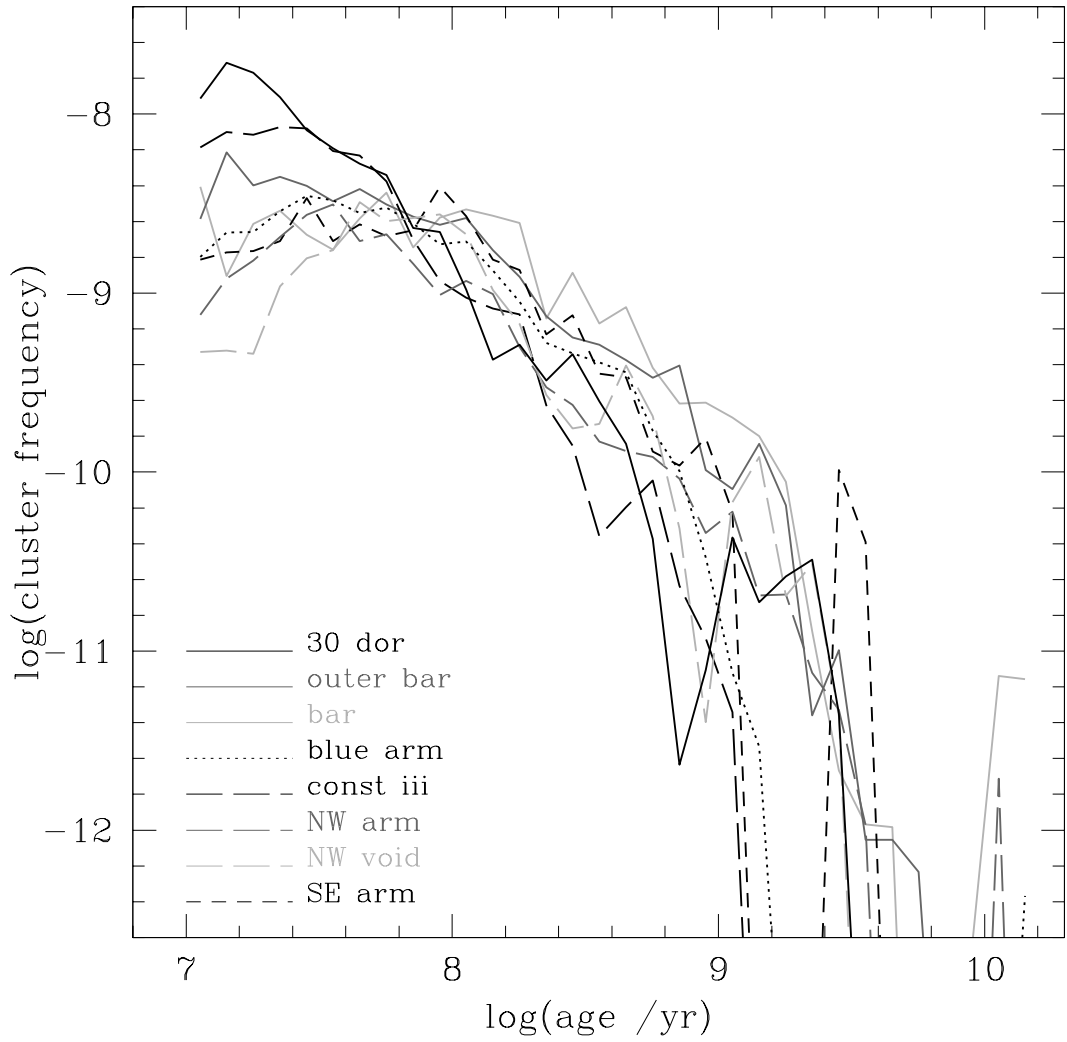,width=144mm}}
\caption{Mass cut-off CFs for different HZ regions.}
\label{fig5}
\end{figure}

\begin{figure}
\centerline{\psfig{figure=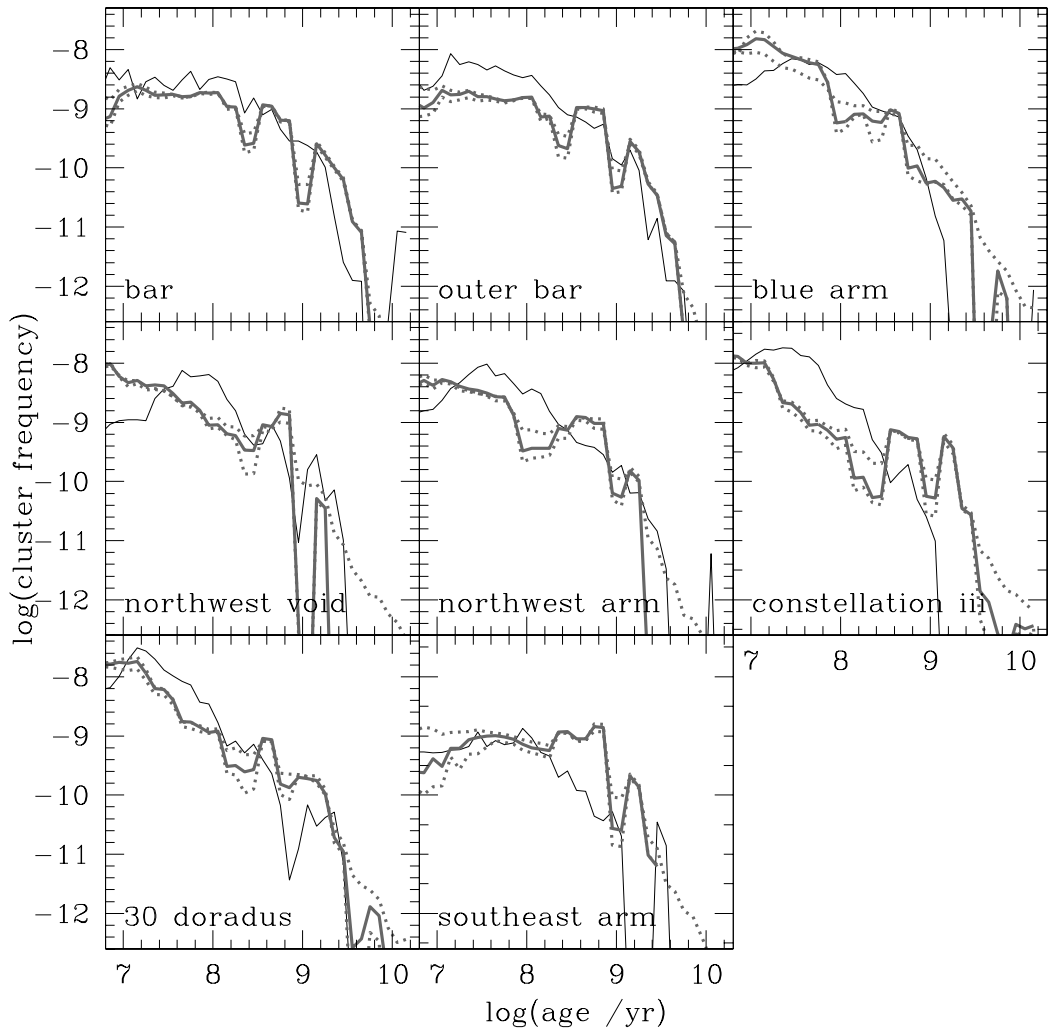,width=144mm}}
\caption{Comparison between the observed CFs (thin solid line) and the theoretical ones
( thick gray solid line). The error curves for the latter are drawn with thick gray dotted lines.}
\label{fig6}
\end{figure}

\clearpage

\appendix

\section{Completeness in the cluster frequencies}

We provide here with a series of experiments related to the effect that the
clusters without age estimates might cause in the derived CFs (see Sect. 3). We produced
multiple panel figures for the eigth HZ regions. Each panel shows the
previously derived CF for the respective HZ region with a solid line, and the
resultant CF according to different age distributions for the clusters without
ages drawn with dotted lines. The employed normalized age distributions are included 
at the bottom of each panel.

\begin{figure}
\centerline{\psfig{figure=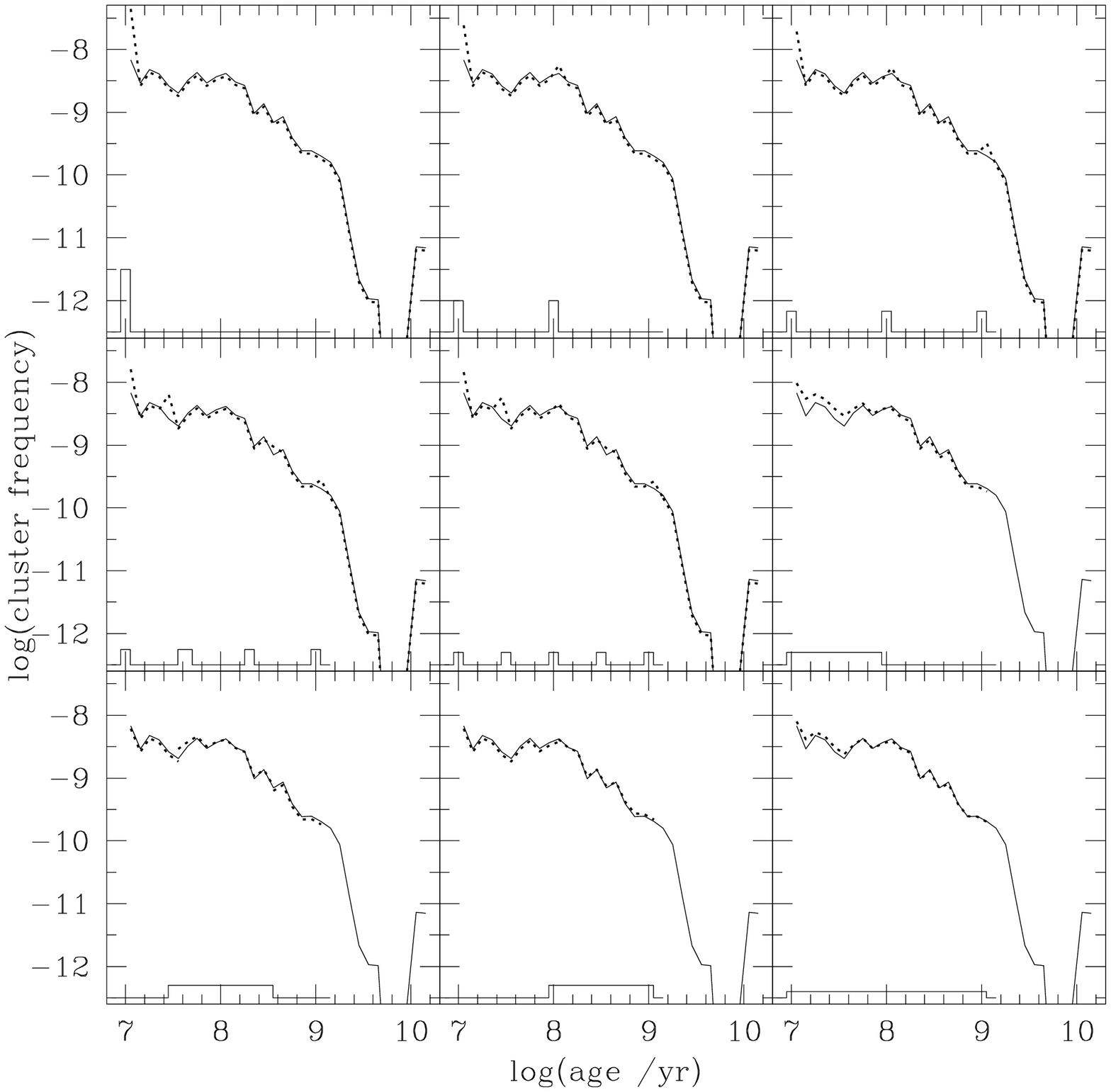,width=144mm}}
\caption{CFs for the LMC Bar (see Appendix A for details).}
\label{figA1}
\end{figure}

\begin{figure}
\centerline{\psfig{figure=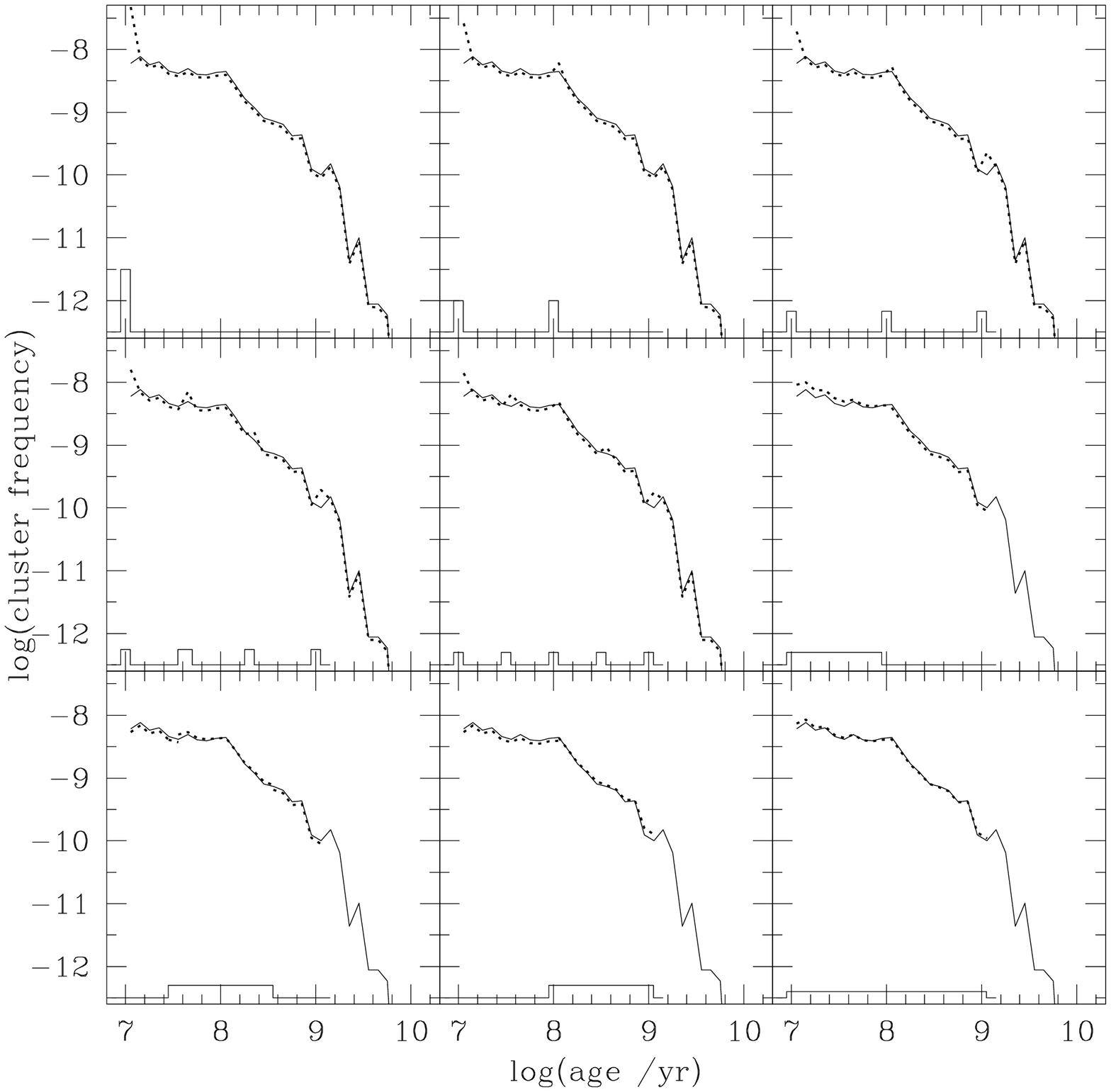,width=144mm}}
\caption{CFs for the LMC Outer Bar (see Appendix A for details).}
\label{figA2}
\end{figure}

\begin{figure}
\centerline{\psfig{figure=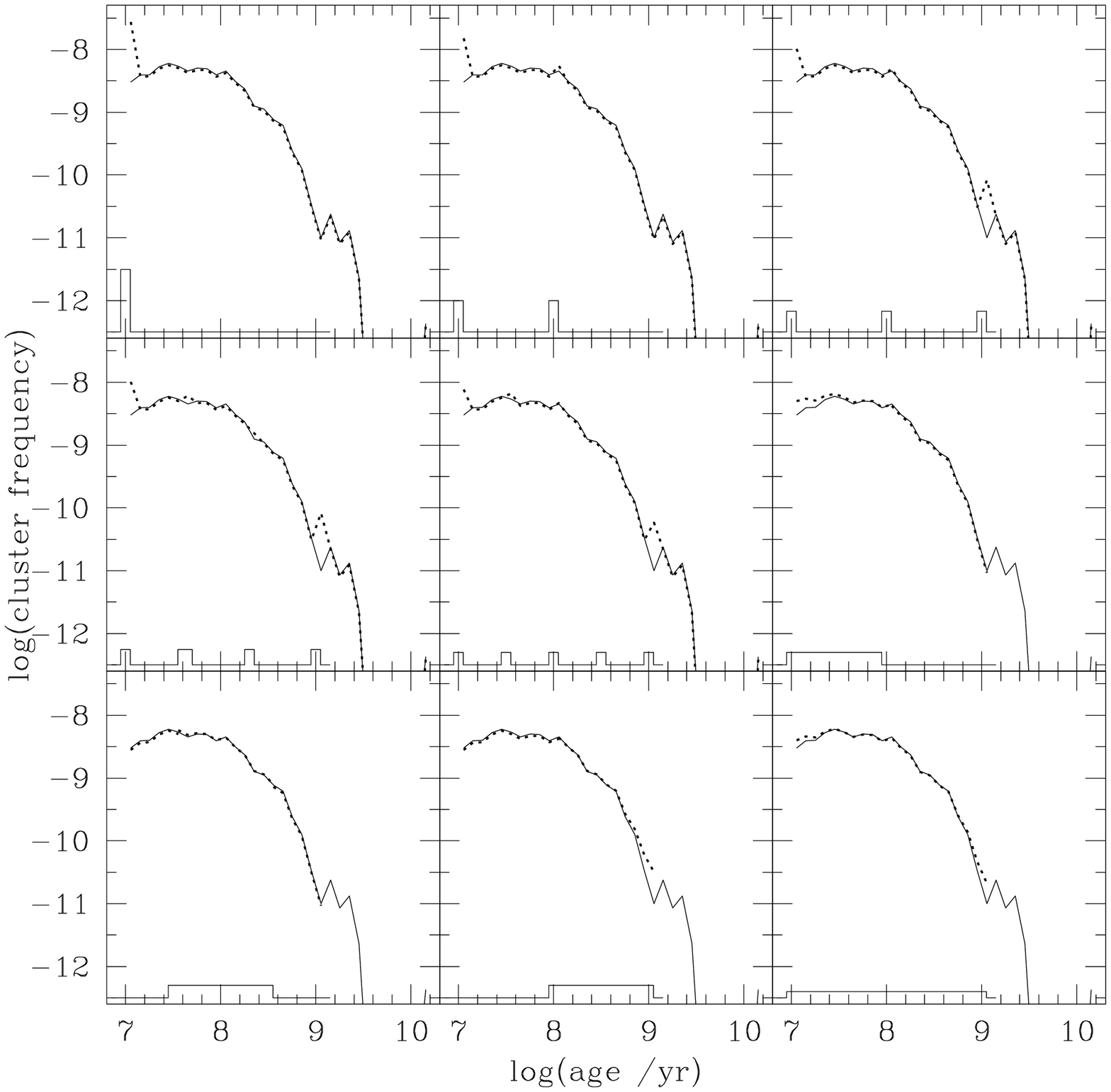,width=144mm}}
\caption{CFs for the LMC Blue Arm (see Appendix A for details).}
\label{figA3}
\end{figure}

\begin{figure}
\centerline{\psfig{figure=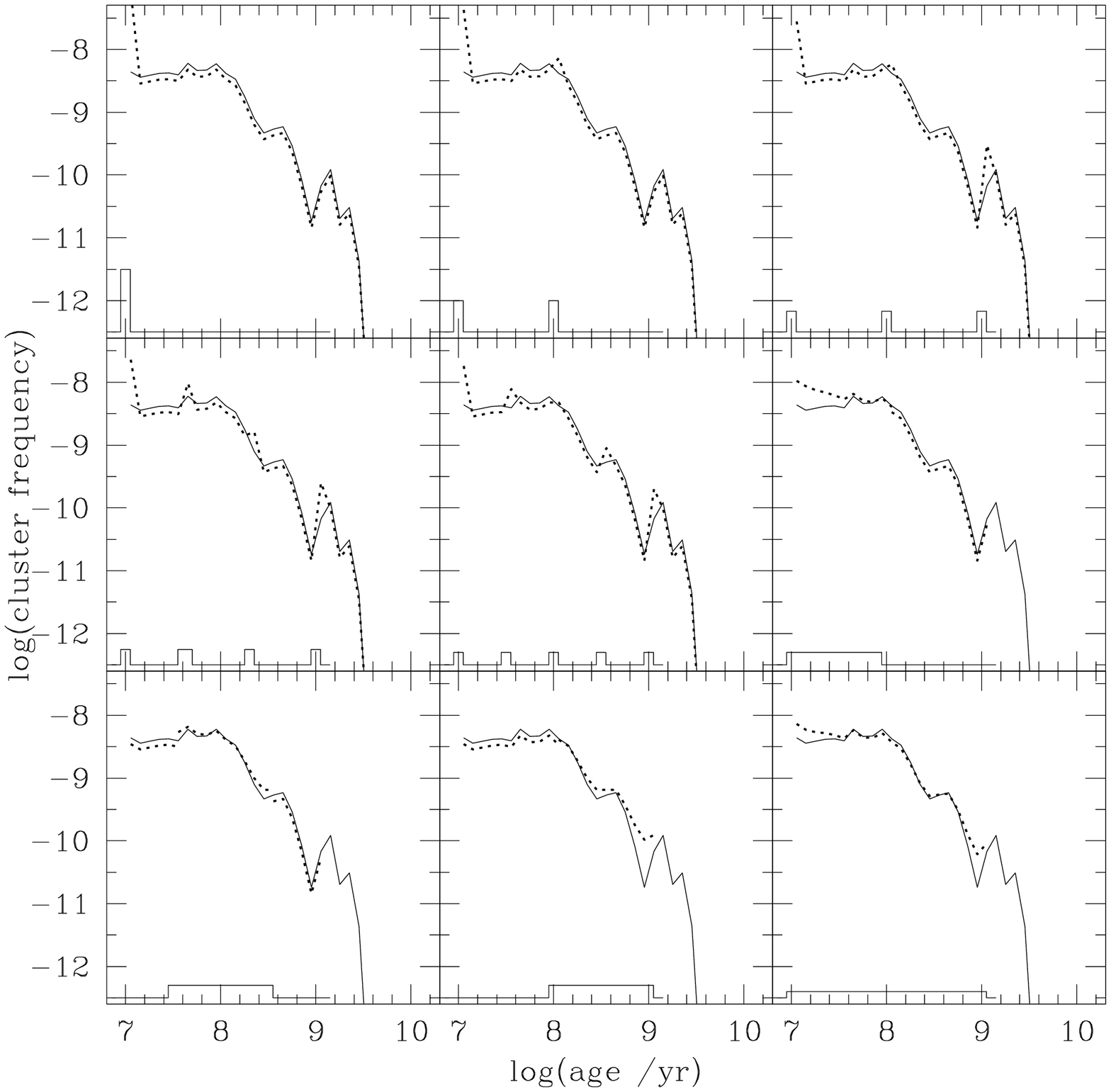,width=144mm}}
\caption{CFs for the LMC Northwest Void (see Appendix A for details).}
\label{figA4}
\end{figure}

\begin{figure}
\centerline{\psfig{figure=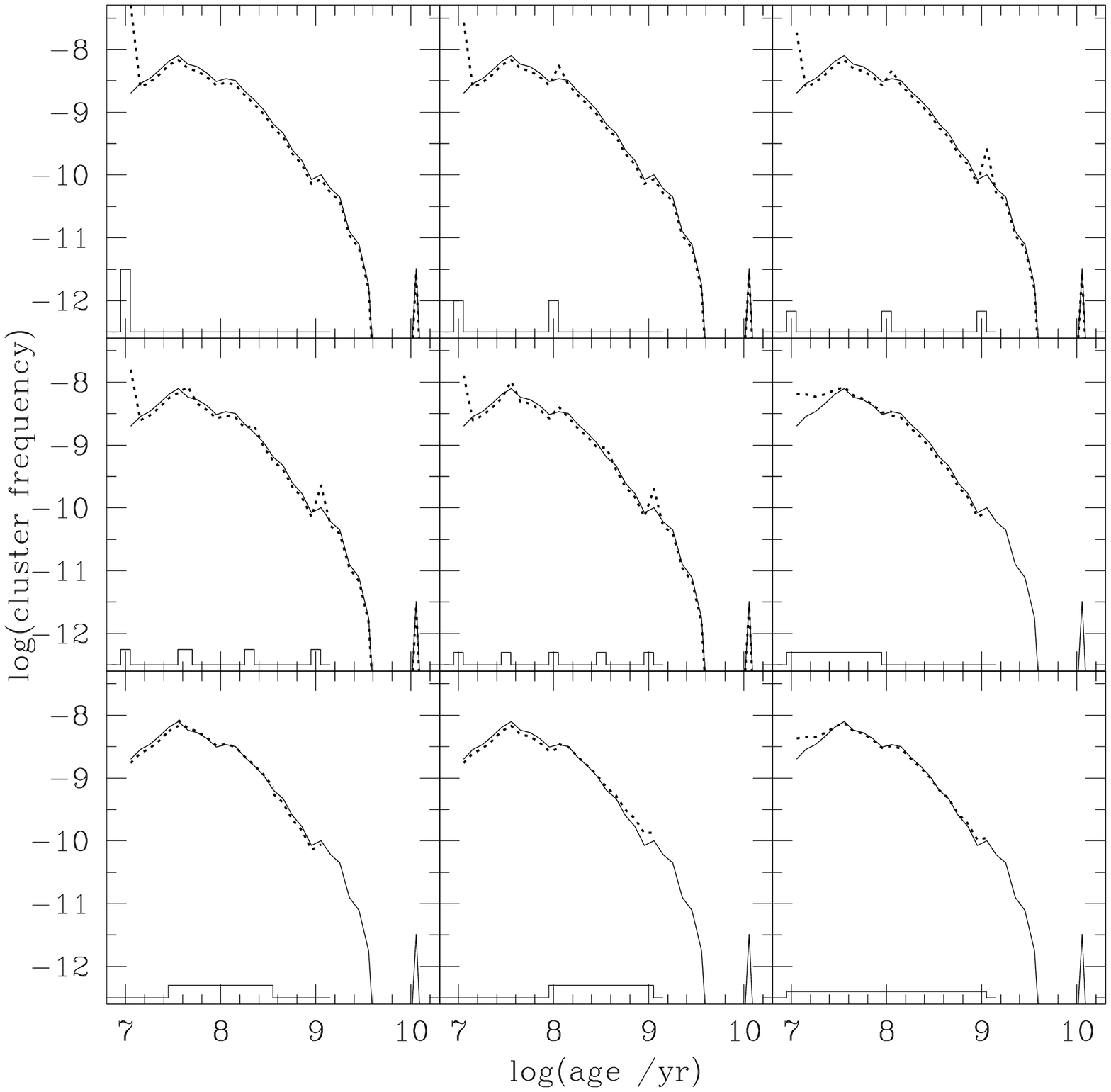,width=144mm}}
\caption{CFs for the LMC Northwest Arm (see Appendix A for details).}
\label{figA5}
\end{figure}

\begin{figure}
\centerline{\psfig{figure=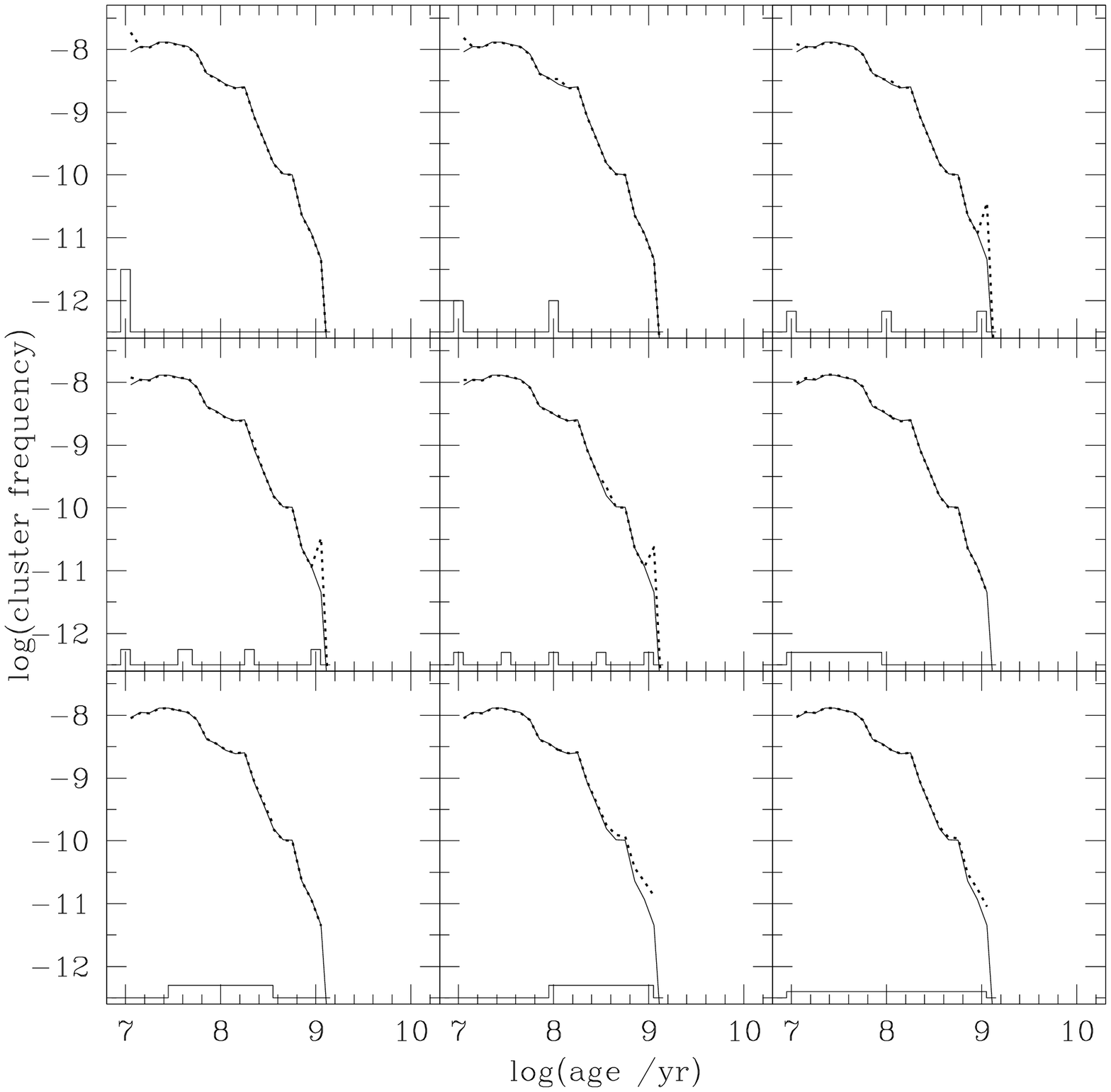,width=144mm}}
\caption{CFs for the LMC Constellation III (see Appendix A for details).}
\label{figA6}
\end{figure}

\begin{figure}
\centerline{\psfig{figure=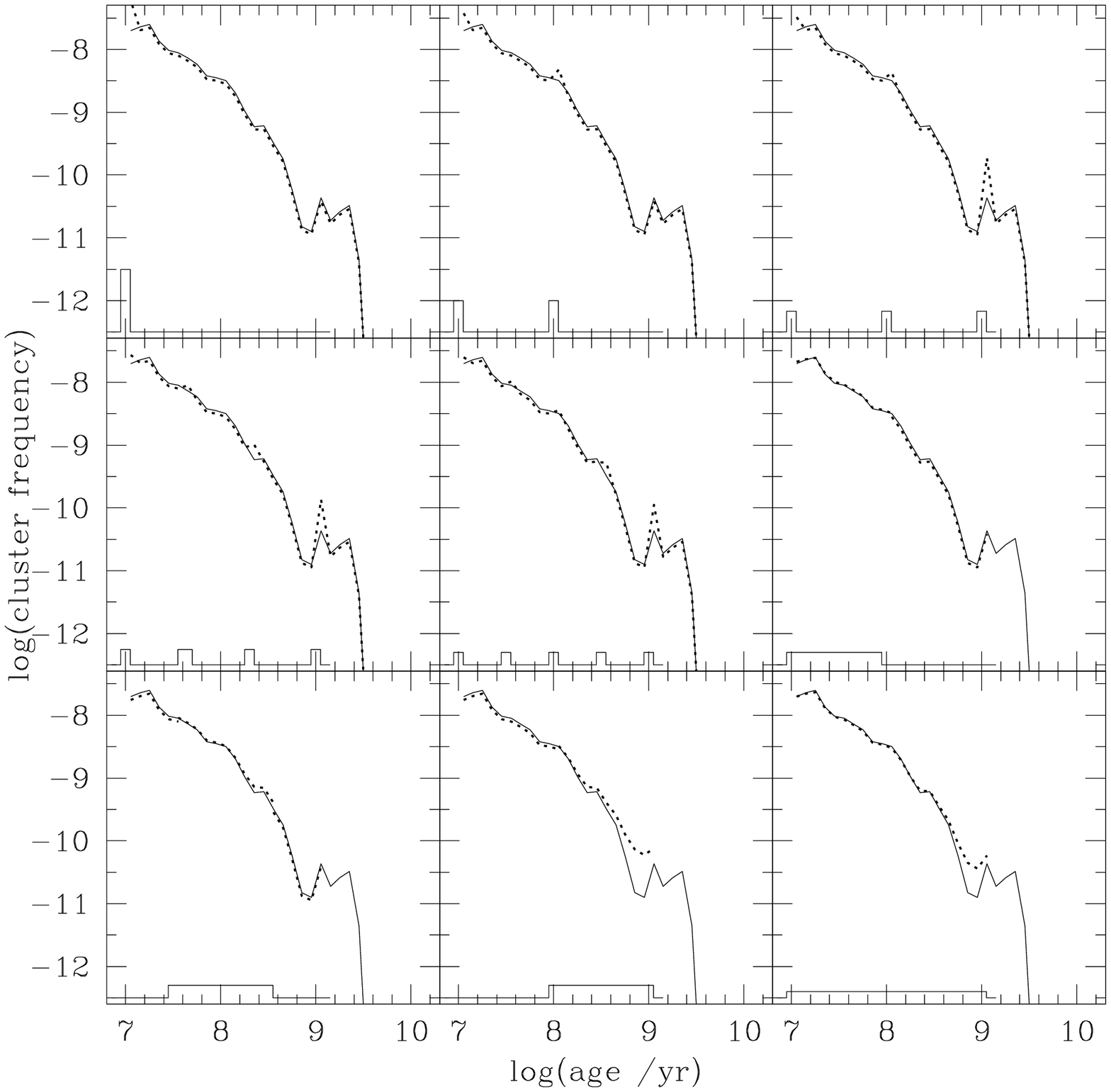,width=144mm}}
\caption{CFs for the 30 Doradus (see Appendix A for details).}
\label{figA7}
\end{figure}

\begin{figure}
\centerline{\psfig{figure=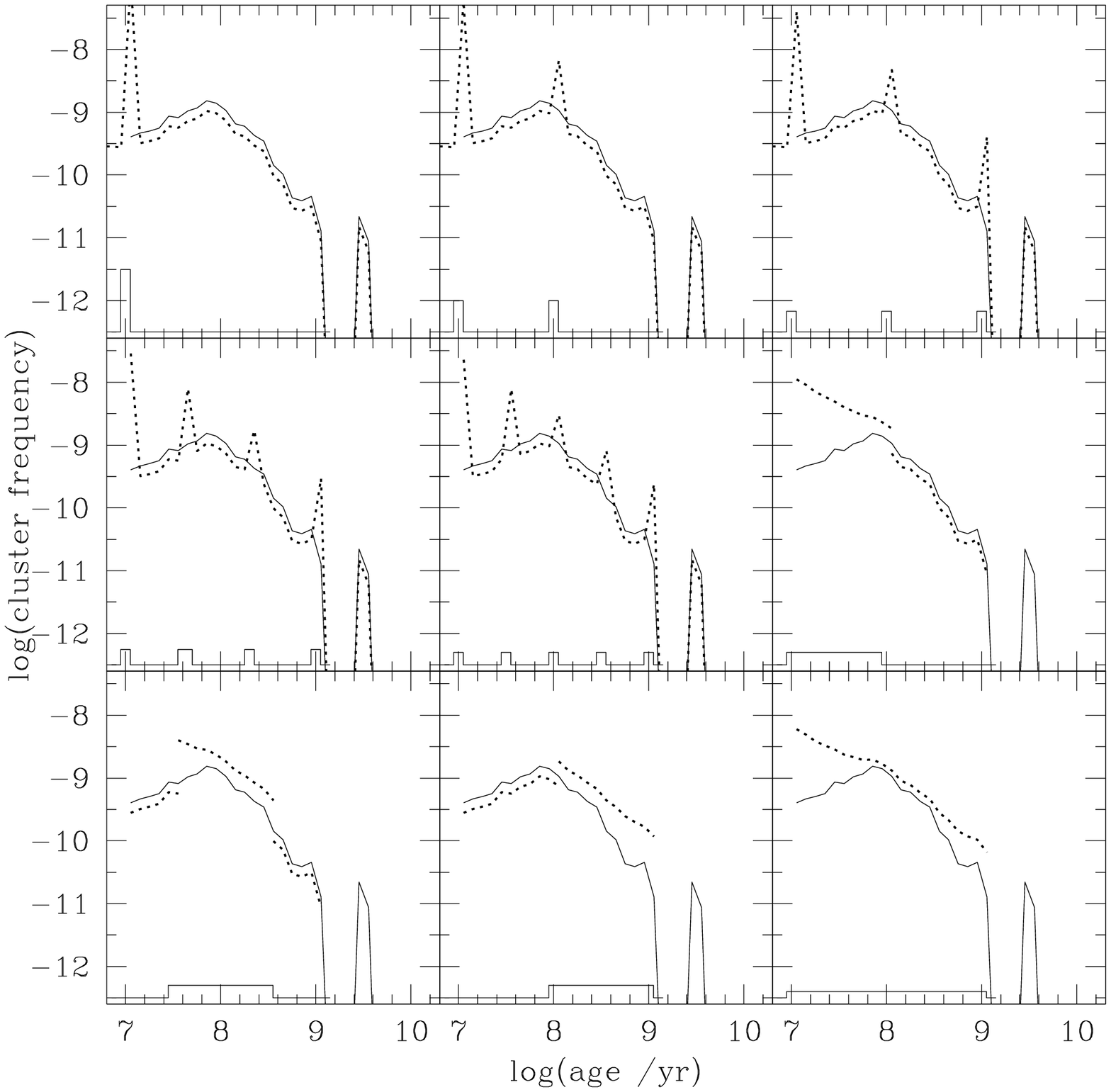,width=144mm}}
\caption{CFs for the LMC Southeast Arm (see Appendix A for details).}
\label{figA8}
\end{figure}

\end{document}